\documentclass[aps,pra,reprint,superscriptaddress,flushbottom,longbibliography]{revtex4-1}

\usepackage{general_latex}

\begin{document}

\title{Optimal and near-optimal probe states for quantum metrology of number conserving two-mode bosonic Hamiltonians}
\author{T.\,J.~Volkoff}
\email{volkoff@snu.ac.kr}
\affiliation{Department of Physics and Astronomy, Center for Theoretical Physics, Seoul National University,   
Seoul 151-747, Korea}

\begin{abstract}
We derive families of optimal and near-optimal probe states for quantum estimation of the coupling constants of a general two-mode number-conserving bosonic Hamiltonian describing one-body and two-body dynamics. We find that the optimal states for estimating the dephasing of the modes, the self-interaction strength, and the contact interaction strength are related to the NOON states, whereas the optimal states for estimation of the intermode single particle tunneling amplitude are superpositions of antipodal SU(2) coherent states. For estimation of the amplitude of pair tunneling and the amplitude of density-dependent single particle tunneling processes, respectively, we introduce classes of variational superposition probe states that provide near perfect saturation of the corresponding quantum Cram\'{e}r-Rao bounds. We show that the ground state of the pair tunneling term in the Hamiltonian has a high fidelity with the optimal states for estimation of a single particle tunneling amplitude, suggesting that high-performance probes for tunneling amplitude estimation may be produced by tuning the two-mode system through a quantum phase transition.  
\end{abstract}
\maketitle

\section{Introduction}

Methods for generating nonclassical states of the electromagnetic field or atomic ensembles by unitary or dissipative quantum dynamics are central to atom-field based quantum technologies, e.g., estimation of dynamical parameters beyond the standard quantum limit.
In the realm of degenerate massive bosons (including, e.g., ultracold bosonic gases), studies on nonclassical states of $N$ interacting bosons distributed among two orthogonal single particle states \cite{wallstwomode,leehuangnonlin,spekkensdwbec} have led to increased understanding of phenomena such as many-body entanglement \cite{jakschzoller}, Schr\"{o}dinger cat state formation \cite{ruostekoski,moorehuang}, and coherent pair tunneling \cite{sahagunpairbosehubb}. These systems also provide a foundation for the modern atomic clock \cite{ludlowye}. Furthermore, systems of two-mode bosons exhibiting various types of tunneling dynamics are useful as models of bosonic quantum phase transitions beyond the well understood insulator-superfluid transition of the Bose-Hubbard model. Therefore, it is of considerable interest to identify and characterize the families of states that allow us to probe the dynamical parameters of systems of two-mode bosons near the ultimate quantum limit imposed by the quantum Cram\'{e}r-Rao (QCR) bound \cite{holevobook}.

The possibility of coherent many-boson processes described by monomials $a_{0}^{\dagger m}a_{1}^{n}+h.c.$ for $m,n >1$ complicates the allowed two-mode dynamics of bosons compared to fermions. However, by analyzing a weakly interacting Bose gas we can restrict to generic one-particle and two-particle processes.
A generic number-conserving Hamiltonian $H$ governing the motion of $N$ weakly interacting bosons takes the form 
\begin{eqnarray}
H&=&\vartheta \left( a_{1}^{\dagger}a_{1} - a_{0}^{\dagger}a_{0}\right) +  \sum_{j,k=0}^{1}V_{jk}a_{k}^{\dagger}a_{k}a_{j}^{\dagger}a_{j} \nonumber \\ &+&   \sum_{k=1}^{2}\left(A_{k} a_{0}^{\dagger k}a_{1}^{k}+\text{H.c.}\right) \nonumber \\ &+&    \sum_{k=0}^{1}\left(T_{k} a_{k}^{\dagger }a_{k}a_{0}^{\dagger}a_{1} + \text{H.c.} \right)
\label{eqn:ham}
\end{eqnarray}
where $V_{jk}$ is a real, symmetric $2 \times 2$ matrix. $H$ is a two-mode or two-site version of the extended Bose-Hubbard model, which has been considered previously in various physical contexts \cite{liangpairtunn,jurgensen,johansson,lewensteinrev}.

In this paper, we identify families of pure quantum probe states that exhibit optimal or near-optimal QCR bounds for single-parameter estimation of each of the coupling constants of the Hamiltonian Eq.(\ref{eqn:ham}). Identification of novel near-optimal probe states serves two purposes in the physics of quantum metrology: 1) it informs the structure of the optimal measurement of the parameter, and 2) it allows to determine the types of particle correlations that are relevant for high-precision estimation of the parameter. The latter information leads to a metrological ``phase diagram'' in which physical characteristics of quantum states are used to indicate their metrological usefulness (i.e., maximal quantum Fisher information).

Two important complementary problems are not treated in the present work: 1) the identification of optimal probe states for simultaneous estimation of multiple parameters of Eq.(\ref{eqn:ham}) for cases in which the QCR bound is achievable, and, 2) specific protocols for experimental generation of the optimal or near-optimal states. However, Sec. \ref{sec:achieving} contains a brief discussion of the problem of simultaneous quantum estimation of two or more coupling constants of Eq.(\ref{eqn:ham}). We  note that proposals exist for the experimental generation of the superpositions of antipodal coherent states \cite{lapertferrini} and NOON-type states \cite{mullinlaloe}, which we show to be optimal for quantum estimation of $A_{1}$ and $\vartheta$ respectively.  In contrast, experimental generation of the variational probe states that allow for near-optimal quantum estimation of the number-weighted tunneling constants $T_{0}$, $T_{1}$, and the pair tunneling constant $A_{2}$, requires methods for generating entanglement between pairs of bosons in addition to methods for generating coherent superposition states of the entire system of particles.

A brief outline of the paper is as follows: Sec. \ref{sec:varback} contains a derivation of the terms appearing in Eq.(\ref{eqn:ham}) from the microscopic theory of the weakly interacting Bose gas and provides the basic setting for variational quantum estimation of a real coupling constant. In Sec. \ref{sec:contact}, we derive the optimal families of states for estimation of the coupling constants of the terms of Eq.(\ref{eqn:ham}) that are diagonal in the basis of Dicke states. In Sec. \ref{sec:tunnel}, we show that equal weight superpositions of antipodal spin-$N/2$ coherent states are optimal probes of the single particle tunneling coupling constants. Sec. \ref{sec:pairtun} contains a brief background on coherent pair tunneling and introduces a class of variational probe states that provides a near-minimal QCR bound for estimation of $A_{2}$. We also show that a high fidelity variational ground state of the pair tunneling Hamiltonian $a_{0}^{\dagger 2}a_{1}^{2}+h.c.$ can be used as a probe for quantum estimation of the single particle tunneling amplitude. In Sec. \ref{sec:weighttun}, we derive near-optimal families of variational probe states for the number-weighted tunneling amplitudes $T_{0}$, $T_{1}$ by similar methods as used in the pair tunneling estimation problem.

\section{Variational quantum metrology for massive two-mode bosons\label{sec:varback}}

\subsection{ Optimal and near-optimal families of states \label{sec:problemstatement}}
The Hamiltonian in Eq.(\ref{eqn:ham}) arises naturally from the dynamics of the weakly interacting Bose gas. Throughout this work, we omit projections that restrict $H$ to the symmetric $N$ boson Hilbert space $S(\mathbb{C}^{2})^{\otimes N}$, where $S$ symmetrizes the tensor product basis. In practice, this amounts to using $a_{0}^{\dagger}\ket{N,0} = a_{1}^{\dagger}\ket{0,N} =0$ or, equivalently, $[a_{0},a_{0}^{\dagger}]=\mathbb{I}-(N+1)\ket{N,0}\bra{N,0}$, $[a_{1},a_{1}^{\dagger}]=\mathbb{I}-(N+1)\ket{0,N}\bra{0,N}$. 
The energy difference between the single particle states $\ket{0}$ and $\ket{1}$, corresponding to single particle wavefunctions $\phi_{0}(x)$, $\phi_{1}(x)$, respectively, is given by $2\vartheta$. Substitution of a generic two-mode field operator  $\hat{\psi}(x) = (1+\vert z \vert^{2})^{-1/2}(\phi_{0}(x)a_{0}+z\phi_{1}(x) a_{1})$ in the weakly interacting Bose gas Hamiltonian $H_{\text{WBG}}:=\int_{\Omega}d^{3}x \, \left( {-\hbar^{2} \over 2m}\hat{\psi}^{\dagger}\nabla^{2}\hat{\psi} + {V_{0}\over 2}(\hat{\psi}^{\dagger}\hat{\psi})^{2} \right)$, with $\Omega$ a compact subset of $\mathbb{R}^{3}$, allows to make the following identifications with the underlying parameters in Eq.(\ref{eqn:ham}):
\begin{eqnarray}
V_{00}&=&g(\vert z\vert)\langle \vert \phi_{0} \vert^{2},\vert \phi_{0} \vert^{2} \rangle_{2} \nonumber \\
V_{11}&=&\vert z \vert^{4}g(\vert z\vert)\langle \vert \phi_{1} \vert^{2},\vert \phi_{1} \vert^{2} \rangle_{2} \nonumber \\
V_{01}&=&4\vert z\vert^{2}g(\vert z\vert)\langle \vert \phi_{0} \vert^{2},\vert \phi_{1} \vert^{2} \rangle_{2} \nonumber \\
A_{2} &=& z^{2}g(\vert z\vert)\langle  \phi_{0}^{2},\phi_{1}^{2} \rangle_{2} \nonumber \\ 
T_{0} &=& zg(\vert z\vert)\langle \vert \phi_{0}\vert^{2} \phi_{0},\phi_{1} \rangle_{2} \nonumber \\ 
T_{1} &=& z\vert z \vert^{2}g(\vert z\vert)\langle  \phi_{0},\phi_{1} \vert \phi_{1}\vert^{2}\rangle_{2},
\label{eqn:identify}
\end{eqnarray}

where $g(\vert z\vert) := (V_{0}/2)(1+\vert z \vert^{2})^{-2}$ and $\langle f , h \rangle_{2} := \int_{\Omega} d^{3}x \overline{f(x)}h(x)$ is the  inner product on $L^{2}(\Omega)$ \footnote{Note that for the three mode weakly interacting Bose gas in the Bogoliubov c-number substitution approximation for the lowest mode (i.e., the atom laser approximation), there arise parametric amplification and downconversion terms. We do not consider these number non-conserving processes here.}. The phase difference $\vartheta$ and tunneling amplitude $A_{1}$ are found similarly from the kinetic energy term of $H_{\text{WBG}}$.  Physical interpretations of the coupling constants of Eq.(\ref{eqn:identify}) are as follows: $V_{00}$ ($V_{11}$) represents the interaction energy of particles in the $\ket{0}$ ($\ket{1}$) state, $V_{01}$ represents the intermode interaction energy (e.g., if $\ket{0}$ and $\ket{1}$ are spatially localized modes, then it represents the energy of the contact interaction), $A_{2}$ is the tunneling amplitude for coherent tunneling of pairs of particles between the $\ket{0}$ and $\ket{1}$ modes, and $T_{0}$ and $T_{1}$ are the density-dependent single particle tunneling amplitudes. In the weakly interacting lattice Bose gas in the nearest-neighbor approximation, the coupling constants $A_{2}$ and $V_{01}$ are significantly smaller than the other coupling constants \cite{lewensteinrev}. In contrast, in dipolar lattice Bose gases, the contact interaction $V_{01}$ and density-dependent tunneling amplitudes $T_{0}$, $T_{1}$ are non-negligible and can dominate the dynamics \cite{lewensteindens}. Proposals exist for engineering the pair tunneling amplitude $A_{2}$ in the weakly interacting Bose gas by introducing a time-dependent interaction strength \cite{opatrnylipkin}. Keeping an eye toward quantum metrology, Eq.(\ref{eqn:ham}) contains a minimal set of terms such that the combined information obtained from optimal estimation of the coupling constant of each term allows the most precise description of the dynamics possible.

For a time-independent observable $A$ generating a time evolution $U(t)=e^{-it\kappa A / \hbar}$, where $\kappa$ is a real coupling constant, the QCR bound for the variance of an unbiased estimator $\hat{\kappa}$ of $\kappa$ is \cite{holevobook}
\begin{equation}
\langle (\Delta \hat{\kappa})^{2}\rangle \ge {\hbar^{2} \over \nu T^{2}\mathcal{F}(\rho )}.
\label{eqn:qcrtime}
\end{equation}
In Eq.(\ref{eqn:qcrtime}), $\nu$ is the number of probe states $\rho$ consumed in the estimation protocol (i.e., the number of experiments run), $\mathcal{F}(\rho)$ is the quantum Fisher information (QFI) on the unitary path generated by $A$, and $T$ is a time resource that is externally or internally calibrated. In this work, we do not consider the errors associated with the determination of either $\hbar$ or $T$ and absorb the factor $T/\hbar$ into the measurement of the estimator $\hat{\kappa}$.
With this convention of units, the maximum value of the QFI for a self-adjoint $A$ is $(\lambda_{\text{max}} -\lambda_{\text{min}})^2$, where $\lambda_{\text{max(min)}}$ is the maximal (minimal) eigenvalue of $A$. If $\lambda_{\text{max}}$ and $\lambda_{\text{min}}$ each have geometric multiplicity of 1, then the maximal QFI is obtained for probe states belonging to the family of superpositions $\mathcal{S}_{A}:= \lbrace 1/\sqrt{2} \left( \ket{\lambda_{\text{min}}} + e^{i\eta}\ket{\lambda_{\text{max}}} \right) : \, \eta \in [0,2\pi) \rbrace$. Considered as a set in projective Hilbert space, $\mathcal{S}_{A}$ is stabilized by the time-evolution generated by $A$, i.e., $e^{-itA}\mathcal{S}_{A} = \mathcal{S}_{A}$. In Secs. \ref{sec:pairtun} and \ref{sec:weighttun}, we consider cases for which $A$ exhibits a chiral symmetry, i.e., there exists a unitary $U$ such that $UAU^{\dagger} = -A$. In this case, $\lambda_{\text{min}} = -\lambda_{\text{max}}$ and the optimal family of superpositions are those eigenvectors of $A^{2}$ with eigenvalue $\lambda_{\text{max}}^{2} = \lambda_{\text{min}}^{2}$ such that $\langle A \rangle = 0$. We note that equality in the QCR bound for estimation of the real parameter $\theta$ of the path $e^{-i\theta A}$ with probe state $\ket{\Psi}$ can be achieved by a projection-valued measurement with measurement elements consisting of projectors onto the eigenvectors of the symmetric logarithmic derivative corresponding to $A$ and $\ket{\Psi}$ (see section \ref{sec:achieving}) \cite{helstrombook}.

When maximal and minimal eigenvectors $\ket{\lambda_{\text{max}(\text{min})}}$ of the generating observable $A$ are not solvable analytically, one may opt to utilize variational states $\ket{\psi_{\text{max}(\text{min})}}$ corresponding to $\ket{\lambda_{\text{max}(\text{min})}}$, respectively, in order to approximate elements of the optimal family $\mathcal{S}_{A}$. Consider variational states $\ket{\psi_{\text{max}(\text{min})}}$ satisfying  $\langle \psi_{\text{min}} \vert \lambda_{\text{min}} \rangle = \langle \psi_{\text{max}} \vert \lambda_{\text{max}} \rangle = \sqrt{1-\epsilon}$ with $0<\epsilon <1$ and assume the following two conditions: 1) $\langle \psi_{\text{min}} \vert \lambda_{\text{max}} \rangle = \langle \psi_{\text{max}} \vert \lambda_{\text{min}} \rangle = 0$, and 2) $\langle \psi_{\text{min}} \vert \psi_{\text{max}} \rangle = w$ with $0\le \vert w\vert \le \sqrt{\epsilon} $. The upper bound imposed on $\vert w\vert$ is derived by considering the possibility that the Born probabilities for $\ket{\psi_{\min}}$ in the subspaces $\mathbb{C}\ket{\lambda_{\text{min}}}$ and $\mathbb{C}\ket{\psi_{\text{max}}}$ sum to 1. We have taken $\langle \psi_{\text{min}} \vert \lambda_{\text{min}} \rangle = \langle \psi_{\text{max}} \vert \lambda_{\text{max}} \rangle = \sqrt{1-\epsilon}$ as an assumption because the coupling constants in the Hamiltonian Eq.(\ref{eqn:ham}) that necessitate use of a variational probe state are associated to self-adjoint operators that have their spectrum symmetric about 0. In particular, given $\ket{\psi_{\text{min}}}$ such that $\langle \psi_{\text{min}} \vert \lambda_{\text{min}} \rangle =\sqrt{1-\epsilon}$, application of the symmetry operation to $\ket{\psi_{\text{min}}}$ generates a variational approximation $\ket{\psi_{\text{max}}}$ to the highest eigenvector $\ket{\lambda_{\text{max}}}$ such that $ \langle \psi_{\text{max}} \vert \lambda_{\text{max}} \rangle = \sqrt{1-\epsilon}$.  Under these conditions, an element of $\mathcal{S}_{A}$ defined by the relative phase $\eta$ exhibits the largest fidelity with the following state in the two dimensional Hilbert space spanned by $\ket{\psi_{\text{min}}}$ and $\ket{\psi_{\text{max}}}$:
\begin{equation}
 {(1-we^{i\eta})\ket{\psi_{\text{min}}} + (e^{i\eta}-w)\ket{\psi_{\text{max}}} \over \sqrt{2(1-w^{2})(1-w\cos \eta)}}.
 \label{eqn:variationalgen}
\end{equation}
The inner product of this state with its corresponding element of $\mathcal{S}_{A}$ is $\sqrt{1-\epsilon}\sqrt{(1-w\cos\eta) / (1-w^{2})}$.

Consider the problem of quantum estimation of the real parameter $\theta$ defining the unitary path $e^{-i\theta A}$, where $A$ is bounded, self-adjoint, and has spectrum symmetric about zero. A well-defined criterion for a variational superposition state $\ket{\psi_{\text{var}}}$ approximating an element of the true optimal family $\ket{\psi_{\text{true}}} \in \mathcal{S}_{A}$ to be useful is that it satisfy $1-\vert \langle \psi_{\text{true}} \vert \psi_{\text{var}}\rangle \vert^{2} \in \mathcal{O}(\Vert A \Vert^{-2})$.  This criterion can be understood by noting that if $\langle \psi_{\text{var}} \vert A \vert \psi_{\text{var}} \rangle = 0$, i.e., the variational probe has energy zero with respect to Hamiltonian $A$, the inequality
\begin{equation}
 \langle (\Delta A)^{2} \rangle_{\ket{\psi_{\text{true}}}}-\langle (\Delta A)^{2} \rangle_{\ket{\psi_{\text{var}}}}  \le \Vert A\Vert^{2}\left(1-\vert \langle \psi_{\text{true}} \vert \psi_{\text{var}}\rangle \vert^{2} \right)
 \label{eqn:varineq}
\end{equation}
holds.  Equation (\ref{eqn:varineq}) shows that the requirement $1-\vert \langle \psi_{\text{true}} \vert \psi_{\text{var}}\rangle \vert^{2} \in \mathcal{O}(\Vert A \Vert^{-2})$ is equivalent to a $\mathcal{O}(1)$ maximal difference in QFI between the variational and optimal families.

\subsection{Achieving the QCR bound\label{sec:achieving}}
A single-shot quantum metrology protocol  (i.e., possessing a QCR bound with $\nu = 1$ in Eq.(\ref{eqn:qcrtime})) can be divided into the following four steps: 1) high-fidelity preparation of the probe system in the desired probe quantum state, 2) parametrized dynamics applied to the probe state, and 3) measurement of an observable corresponding to an estimator of the parameters, and 4) classical post-processing of the measurement results. When a pure state probe $\ket{\Psi}$ is utilized in a protocol for estimation of the single real parameter $\theta$ defining a one-parameter path $e^{-i\theta A}$, $\ket{\Psi}$ is imprinted with the path parameter via $\ket{\Psi_{\theta}}:=e^{-i\theta A}\ket{\Psi}$. Then, the symmetric logarithmic derivative operator $L_{\theta}$ defined by \begin{equation} \del_{\theta}\ket{\Psi_{\theta}}\bra{\Psi_{\theta}} = (1/2)(L_{\theta}\ket{\Psi_{\theta}}\bra{\Psi_{\theta}} + \ket{\Psi_{\theta}}\bra{\Psi_{\theta}} L_{\theta})\end{equation} is given by $L_{\theta} = 2i[\ket{\Psi_{\theta}}\bra{\Psi_{\theta}},A]=L_{\theta}^{\dagger}$   \cite{helstrombook}. If $\ket{\Psi}$ is not an eigenvector of $A$, then $L_{\theta}$ is observable with matrix rank 2. The two eigenvectors of $L_{\theta = 0}$ define the projection-valued measurement that saturates the QCR bound for the probe state $\ket{\Psi}$ \cite{parisquantest}.  Note that an unbiased measurement which allows an optimal estimation of $\theta$ for a probe state $\rho_{1}$ can be a biased measurement of $\theta$ if a different probe state $\rho_{2}$ is used. In the present work, we derive families of optimal probe states for independent quantum metrology protocols, where each protocol produces an estimate of a single, real coupling constant in Eq.(\ref{eqn:ham}). The QCR bounds associated to these protocols are, therefore, separately achievable by optimal measurements.

In a more general setting, a multiparameter quantum metrology problem for the two-mode weakly interacting Bose gas model in Eq.(\ref{eqn:ham}) involves simultaneous estimation of $s$ real parameters, where $1<s\le 12$. The 12 real parameters that must be estimated in a full quantum metrology protocol are comprised of: three real parameters corresponding to the $\mathfrak{su}(2)$ generators, one real parameter defining the contact interaction, two real parameters defining the intraspecies scattering, one complex parameter for pair tunneling, and two complex parameters corresponding to the two types of number-weighted tunneling. In order to achieve equality in the multiparameter QCR bound \cite{holevobook} for estimation of the $n$-tuple of parameters ${\bm \theta} = (\theta_{1},\ldots , \theta_{n})$  when a pure state $\ket{\Psi}$ is used as a probe, it is necessary and sufficient that the Gram matrix of the set of vectors $\lbrace \ket{\ell_{j}} := L_{\theta_{j}}\ket{\Psi} \rbrace$ have real entries \cite{fujiwaraspin}. When the parametrized dynamics are defined by $\ket{\Psi_{\bm \theta}} = \exp ( -i\sum_{j=1}^{n}\theta_{j}A_{j} )$, with $A_{j}=A_{j}^{\dagger}$, this condition is equivalent to $\langle \Psi \vert [A_{i},A_{j}] \vert \Psi \rangle = 0$ for each $(i,j)$ pair. Pure probe states that saturate the QCR bound for simultaneous estimation of three real parameters $\lbrace \alpha_{j}\rbrace_{j=1,2,3}$ of the Hamiltonian $H_{\text{spin}}:= \sum_{j=1}^{3}\alpha_{j}J_{j}$, where $\lbrace J_{j}\rbrace_{j=1,2,3}$ are the generators of a spin-$N/2$ representation of SU(2), were produced in Ref.\cite{datta}. These results are directly applicable to the problem of simultaneous estimation of $\vartheta$ and $A_{0}$ in Eq.(\ref{eqn:ham}). In the setting of simultaneous estimation of the coupling constants of quartic interactions of the two-mode weakly interacting Bose gas, we are not aware of a general method for construction of pure probe states that satisfy the following conditions: 1) allow one to achieve equality in the multiparameter QCR bound, and 2) exhibit $O(N^{4})$ scaling of the diagonal elements of the corresponding QFI matrix.

\section{\label{sec:contact}Self-interactions and contact interactions}

Because the self-interaction and contact interaction terms of $H$ are diagonal in the Dicke state basis $\lbrace \ket{N-k,k} \rbrace_{k=0,\ldots , N}$, it is a straightforward task to derive the family of states that maximizes the variance of each of these terms. For example, the observables $(a_{j}^{\dagger}a_{j})^{2}$, $j=1,2$, exhibit maximal variance of $N^{2}$ in the family of states given by the superpositions
\begin{equation}
\ket{\Psi_{V}} := {\ket{0,N} + e^{i\phi}\ket{N,0}  \over \sqrt{2}}
\label{eqn:noon}
\end{equation}
with $\phi \in [0,2\pi)$, i.e., $\ket{\Psi_{V}}$ are the well-known NOON states that appear in the context of attractive ultracold Bose gases (e.g., $^{7}$Li) confined in a double-well potential \cite{ueda,hocats}.

The observable $a_{1}^{\dagger}a_{1}a_{0}^{\dagger}a_{0}$ has, for $N$ even, maximal eigenvalue $N^{2}/4$ corresponding to state $\ket{{N\over 2},{N\over 2}}$ and, for $N$ odd, maximal eigenvalue $(N^{2}-1)/4$ corresponding to the two-dimensional subspace spanned by the Dicke states $\ket{(N-1)/2,(N+1)/2}$ and $\ket{(N+1)/2,(N-1)/2}$. For both even and odd $N$, the minimal eigenvalue is $0$ corresponding to the family of states $\ket{\Psi_{\phi,\theta}}$ given by
\begin{equation}
\ket{\Psi_{\theta,\phi}} := {\cos({\theta \over 2})\ket{0,N} + \sin({\theta \over 2})e^{i\phi}\ket{N,0}  \over \sqrt{2}}.
\label{eqn:noonalt}
\end{equation} Therefore, for $N$ even, the family of states exhibiting maximal variance of $a_{1}^{\dagger}a_{1}a_{0}^{\dagger}a_{0}$ are the following states $\ket{\Psi_{V_{01}}}$ parametrized by $S^{2}\times S^{1}$: \begin{equation}
\ket{\Psi_{V_{01}}} := {1\over \sqrt{2}}\left( \big\vert {N\over 2},{N\over 2} \big\rangle + e^{i\eta} \ket{\Psi_{\theta,\phi}}\right).
\label{eqn:superposcontact}
\end{equation}
For $N$ odd, the analogous states are parametrized by  $S^{2}\times S^{2} \times S^{1}$.

Whereas the optimal states for estimation of the self-interaction and contact interaction strengths are entangled, it is known that product states of distinguishable particles can exhibit QCR bounds that scale as $\mathcal{O}(1/N^{k})$ with $k>1/2$, i.e., below the standard quantum limit, for estimation of nonlinear coupling strengths \cite{beyondheisenprodstate,dattashaji}. For indistinguishable bosons, it has been shown that states exhibiting vanishing mode entanglement are useful for achieving sub-shot noise sensitivities for estimation of matter wave beamsplitter parameters \cite{subshotwithoutentang}. Therefore, if one aims only to surpass classical metrological limits, it is not necessary to generate the large entanglement and coherence exhibited by the optimal families of states.

\section{\label{sec:tunnel}Single particle tunneling and phase estimation}

The Schwinger boson mapping $J_{x} = {1\over 2}(a_{0}^{\dagger}a_{1}+a_{1}^{\dagger}a_{0} )$, $J_{y} ={1\over 2}( ia_{0}^{\dagger}a_{1}-ia_{1}^{\dagger}a_{0} )$, $J_{z}={1\over 2}(a_{1}^{\dagger}a_{1}-a_{0}^{\dagger}a_{0} )$ of the $\mathfrak{su}(2)$ Lie algebra specified by $[J_{j},J_{k}]=i\epsilon^{jk\ell}J_{\ell}$  allows for a simplification of the problem of finding optimal states for estimation of the coupling constants $\vartheta$ and $A_{1}$ of the single particle terms in Eq.(\ref{eqn:ham}). In particular, by restricting $J_{k}$ to $S(\mathbb{C}^{2})^{\otimes N}$, the task becomes equivalent to the optimal estimation of rotation angles for a single spin-$N/2$ particle. In this section, we show that the optimal states for estimating a rotation about unit vector $\vec{n} = (\sin\theta \cos \varphi , \sin \theta \sin \varphi , \cos \theta)$ generated by $\vec{n}\cdot \vec{J}$  are given by equal weight superpositions of antipodal spin-$N/2$ coherent states.
Generation of superpositions of orthogonal spin coherent states by one-axis twisting of a spin coherent state (e.g., by the Bose-Hubbard interaction between bosons in a double-well potential) were considered in Ref.\cite{pezzetwisting}. Superpositions of spin coherent states generated by the interaction of a single qubit and a spin-1/2 ensemble for a specific time interval $t_{0}$ were shown in Ref.\cite{spillercat} to allow magnetometry below the standard quantum limit.

The optimal states for estimation of the real parameters $\vartheta$ and $A_{1}$ appearing in $H$ (representing, respectively, the dephasing parameter and the single particle tunneling amplitude) can be calculated directly from the lowest and highest energy eigenvectors of $J_{z}$ and $J_{x}$, respectively. To derive these optimal states we make use of the spin-$N/2$ coherent state \cite{gilmoreatomcoherent,karassiov} \begin{eqnarray} \ket{\zeta(\vec{n})} &:=&{ \left( \cos({\theta \over 2})a_{0}^{\dagger} + \sin({\theta \over 2})e^{i\varphi}a_{1}^{\dagger} \right)^{N} \over \sqrt{N!} } \ket{0,0} \nonumber \\ &=& (1+\vert \zeta(\vec{n}) \vert^{2})^{-N/2}{\left( a_{0}^{\dagger}+\zeta(\vec{n}) a_{1}^{\dagger} \right)^{N} \over \sqrt{N!}}\ket{0,0} \nonumber \\ &=&(1+\vert \zeta(\vec{n}) \vert^{2})^{-N/2}e^{\zeta(\vec{n}) J_{+}}\ket{N,0}
\label{eqn:cohstate}
\end{eqnarray}
where $\ket{N-j,j}$ denotes the Dicke state of $N$ bosons with $j$ bosons occupying single particle state $\ket{1}$, $\zeta(\vec{n}) = \tan(\theta / 2)e^{i\varphi} \in \mathbb{C}$ is the stereographic projection (from the south pole of $S^{2}$) of $\vec{n}$, and $J_{+}:=J_{x}+iJ_{y} = a_{1}^{\dagger}a_{0}$ is the raising operator. Note that $\ket{N,0}$ is the ground state of $J_{z}$ and that $J_{+}$ annihilates $\ket{0,N}$. $\ket{0,N}$ corresponds to $\vec{n} = (0,0,-1)$ and $\ket{N,0}$ corresponds to $\vec{n} = (0,0,1)$. The state in Eq.(\ref{eqn:cohstate}) parametrizes all possible true Bose-Einstein condensed states of the two-mode system, i.e., states in which all $N$ particles occupy the quantum state $\cos({\theta \over 2})\ket{0} + \sin({\theta \over 2})e^{i\varphi}\ket{1}$. 

\textit{Proposition 1}: The states exhibiting maximal variance of $\vec{n}\cdot \vec{J}$ where $\Vert \vec{n}\Vert =1$ are  superpositions of spin-$N/2$ coherent states of the form
\begin{equation}
 {1\over\sqrt{2}}\left( \ket{-\overline{\zeta(\vec{n}) }}+ e^{i\eta} \ket{\zeta(\vec{n})^{-1}} \right)
\label{eqn:tunopt}
\end{equation}
where $\eta \in [0,2\pi)$.

The Lemma that follows allows for a shorter proof of Proposition 1. Physically, the Lemma reflects the fact that the coherent states can be equivalently defined by raising the lowest spin state or by lowering the highest spin state \cite{perelomov}.

\textit{Lemma 1}: Let $\ket{\zeta(\vec{n})}$ be defined as in Eq.(\ref{eqn:cohstate}). Then \begin{equation}
\ket{\zeta(\vec{n})^{-1}} = (1+\vert \zeta(\vec{n}) \vert^{2})^{-N/2}e^{\zeta J_{-}}\ket{0,N}.
\end{equation}

\textit{Proof of Lemma 1}: From Eq.(\ref{eqn:cohstate}), we have
\begin{widetext}
\begin{eqnarray}
\ket{\zeta(\vec{n})^{-1}} &=& \left( 1+\vert \zeta(\vec{n})\vert^{-2} \right)^{-N/2}\sum_{j=-N/2}^{N/2}\sqrt{{{N}\choose{{N\over 2}+j}}} \zeta(\vec{n})^{-j-{N\over 2}}\ket{{N\over 2}-j,{N\over 2}+j}\nonumber \\
&\cong & \zeta(\vec{n})^{N}\left( 1+\vert \zeta(\vec{n})\vert^{2} \right)^{-N/2}\sum_{j=-N/2}^{N/2}\sqrt{{{N}\choose{{N\over 2}+j}}} \zeta(\vec{n})^{-j-{N\over 2}}\ket{{N\over 2}-j,{N\over 2}+j} \nonumber \\ &=& \left( 1+\vert \zeta(\vec{n})\vert^{2} \right)^{-N/2}\sum_{j=-N/2}^{N/2}\sqrt{{{N}\choose{{N\over 2}+j}}} \zeta(\vec{n})^{-j+{N\over 2}}\ket{{N\over 2}-j,{N\over 2}+j} \nonumber \\ &=& \left( 1+\vert \zeta(\vec{n})\vert^{2} \right)^{-N/2} \left( \zeta(\vec{n})^{N}\sqrt{{{N}\choose{0}}} \ket{N,0} + \zeta(\vec{n})^{N-1}\sqrt{{{N}\choose{1}}} \ket{N-1,1} + \ldots + \zeta(\vec{n})^{0}\sqrt{{{N}\choose{N}}} \ket{0,N} \right)
\end{eqnarray}
\end{widetext}
where in the second line, ``$\cong$'' denotes that the states are equal in the projective Hilbert space. Making use of the power series expansion of $e^{\zeta J_{-}}$, it is clear that
\begin{equation}
e^{\zeta J_{-}}\ket{0,N}= \sum_{j=0}^{N}\sqrt{{{N}\choose{j}}} \zeta^{j}\ket{j,N-j}.
\end{equation}
The Lemma then follows from the symmetry ${{N}\choose{j}} = {{N}\choose{N-j}} $ of the binomial coefficients. $\square$

\textit{Proof of Proposition 1}: The largest (smallest) eigenvalue of $J_{z}$ is $N/2$ ($-N/2$), associated with the eigenvector $\ket{0,N}$ ($\ket{N,0}$). Therefore, the maximal variance of the observable $J_{z}$ is $N^{2}$, occurs for the family of states $\ket{\Psi_{V}}$ given in Eq.(\ref{eqn:noon}). Also, let $\vec{n} := (\sin \theta \cos \varphi , \sin \theta \sin \varphi , \cos \theta)$ and $\vec{w} := \vec{n} \times (0,0,1) / \Vert \vec{n}\times (0,0,1) \Vert  = (\sin \varphi , -\cos \varphi , 0)$. The Baker-Campbell-Hausdorff formula for SU(2) implies  $e^{-i\theta \vec{w}\cdot \vec{J}}\left( \vec{n}\cdot \vec{J} \right) e^{i\theta \vec{w}\cdot \vec{J}} =  J_{z}$. Therefore, $e^{i\theta \vec{w} \cdot \vec{J}}\ket{N,0}$ is the eigenvector of $\vec{n}\cdot \vec{J}$ with eigenvalue $-N/2$.  The Gaussian decomposition in SU(2) \cite{perelomov} implies that $e^{i\theta \vec{w} \cdot \vec{J}}\ket{N,0} = \ket{-\overline{\zeta(\vec{n})}}$ where $\zeta(\vec{n})=\tan ({\theta \over 2})e^{i\varphi}$ as above. Therefore, $\ket{-\overline{\zeta(\vec{n})}}$ is the eigenvector of $\vec{n}\cdot \vec{J}$ with lowest eigenvalue.

Similarly, $e^{i\theta \vec{w} \cdot \vec{J}}\ket{0,N}$ is the eigenvector of $\vec{n}\cdot \vec{J}$ with eigenvalue $N/2$. Another Gaussian decomposition allows to write $e^{i\theta \vec{w} \cdot \vec{J}} = e^{\zeta (\vec{n})J_{-}}e^{\log (1+\vert \zeta(\vec{n})\vert^{2})}e^{-\overline{\zeta(\vec{n})}J_{+}}$, which implies $e^{i\theta \vec{w} \cdot \vec{J}}\ket{0,N} = (1+\vert \zeta(\vec{n}) \vert^{2})^{-N/2}e^{\zeta(\vec{n})J_{-}} \ket{0,N} = \ket{\zeta(n)^{-1}}$ by Lemma 1. The Proposition now follows from the general form of the family of states introduced in Section \ref{sec:varback} which maximizes the variance of a bounded self-adjoint operator. That $\langle -\overline{\zeta(\vec{n})} \vert \zeta(\vec{n})^{-1} \rangle = 0$ is easily verified. $\square$

The family of optimal states in Eq.(\ref{eqn:tunopt}) can be rewritten in a way that reveals their Schr\"{o}dinger cat state structure. For example, this can be seen by expressing Eq.(\ref{eqn:tunopt}) for $\vec{n} = (1,0,0)$ in terms of the first quantized description:
\begin{eqnarray}
\ket{\Psi_{A_{1}}} &=& {1\over \sqrt{2^{N+1}}\sqrt{N!}}\left( \left( a_{0}^{\dagger}+a_{1}^{\dagger}  \right)^{N} + e^{i\eta}\left(a_{0}^{\dagger}-a_{1}^{\dagger}  \right)^{N} \right)\ket{0,0} \nonumber \\ &=& {1\over \sqrt{2}}\left( \ket{+}^{\otimes N} + e^{i\eta}\ket{-}^{\otimes N} \right)
\end{eqnarray}
where $\ket{\pm}$ are eigenvectors of the Pauli operator $\sigma_{x}$ with positive and negative eigenvalue, respectively. We have labeled this particular family of states $\ket{\Psi_{A_{1}}}$ because it is optimal for estimation of the real constant $A_{1}$ appearing in Eq.(\ref{eqn:ham}). These states are special cases of hierarchical Schr\"{o}dinger cat states \cite{volkoffhier}; in particular, they are Greenberger-Horne-Zeilinger (GHZ) states in the eigenbasis of $\sigma_{x}$. The Schr\"{o}dinger cat state structure of the $\ket{\Psi_{A_{1}}}$ can also be clearly visualized by calculating its Husimi $Q$ distribution on the sphere $S^{2}$.

Physically,  Eq.(\ref{eqn:tunopt}) can be viewed as a superposition of two Bose-Einstein condensate states, e.g., for $\ket{\Psi_{A_{1}}}$, one of the product states appearing in the superposition is characterized by all bosons condensed into the state $\propto \ket{0} + \ket{1}$ while the other product state is characterized by all bosons condensed into $\propto \ket{0}-\ket{1}$. However, the presence of the relative phase $e^{i\eta}$ between the superposed Bose-Einstein condensates appearing in $\ket{\Psi_{A_{1}}}$ obscures the distribution of the bosons among the states $\ket{0}$, $\ket{1}$. This distribution can be analyzed by considering the degree of fragmentation $F_{D}$ of the two-mode system \cite{uedatwomode,badertwomode}, which quantifies the extent to which Bose-Einstein condensation occurs in the single particle mode $\ket{0}$ or $\ket{1}$. Given a state $\rho$ of $N>1$ bosons distributed among two single-particle modes $\ket{0}$ and $\ket{1}$, $F_{D}$ is defined by \begin{equation}
F_{D}:= 1-{\vert \lambda_{+} - \lambda_{-} \vert \over N}
\end{equation} where $\lambda_{\pm}$ are the eigenvalues of the one-particle density matrix $\rho^{(1)}_{\mu \nu} = \langle a_{\mu}^{\dagger}a_{\nu} \rangle$, with $\mu$, $\nu \in \lbrace 0 , 1 \rbrace$ in the two-mode approximation. Defining $N_{\text{tot}} := a_{0}^{\dagger}a_{0} + a_{1}^{\dagger}a_{1}$, it is clear that
\begin{equation}
\rho^{(1)}= \langle N_{\text{tot}} \rangle_{\rho}\left( {\mathbb{I}\over 2} + {\vec{v}\cdot \vec{\sigma} \over 2} \right) 
\end{equation} where $\vec{\sigma}$ is the vector of Pauli matrices and $\vec{v} := (2/\langle N_{\text{tot}} \rangle_{\rho})\left( \langle J_{x} \rangle_{\rho} , \langle J_{y} \rangle_{\rho}, -\langle J_{z} \rangle_{\rho} \right)$. Restricting to states $\rho$ such that $\langle N_{\text{tot}} \rangle_{\rho} = N$, one obtains for the degree of fragmentation:
\begin{equation}
F_{D} = 1-{2\sqrt{\langle J_{x}\rangle_{\rho}^{2}+\langle J_{y}\rangle_{\rho}^{2}+\langle J_{z}\rangle_{\rho}^{2}} \over N}.
\end{equation}
Restricting to the family $\ket{\Psi_{A_{1}}}$ that maximizes the variance of $J_{x}$, we find that $\langle J_{x} \rangle  =\langle J_{y} \rangle = 0$. $\langle J_{z} \rangle$ can be found by expanding the states $\ket{\Psi_{A_{1}}}$ in the Dicke state basis
\begin{equation}
\ket{\Psi_{A_{1}}} = {1\over \sqrt{2^{N+1}}} \sum_{j=0}^{N}\sqrt{ {{N}\choose{j}}} \left( 1+e^{-i\eta}(-1)^{j} \right) \ket{N-j , j}.
\end{equation}
The result is (for $N>1$) $\langle \Psi_{A_{1}} \vert J_{z} \vert \Psi_{A_{1}} \rangle = {\cos \eta \over 2^{N-1}} \sum_{j=0}^{N} {{N}\choose{j}} (-1)^{j}(j-{N\over 2}) = 0$. Because of the rotational symmetry, we see that $F_{D}=1$ for the states exhibiting maximal variance of $J_{x}$, $J_{y}$, $J_{z}$. However, there are mixed states that exhibit the same value of $F_{D}$ as a pure state of the form $\ket{\Psi_{A_{1}}}$. Therefore, a probe state satisfying the condition $F_{D} =1$ is necessary, but not sufficient for optimal estimation of $A_{1}$.

\section{\label{sec:pairtun}Pair tunneling} 

\subsection{\label{sec:analytical}Variational metrologically useful states}
The pair tunneling term $a_{0}^{\dagger 2}a_{1}^{2} + h.c.$ appearing in Eq.(\ref{eqn:ham}) allows boson pairs to coherently tunnel between the $\ket{0}$ and $\ket{1}$ state. For $T_{0} = T_{1} = 0$, the quantum phase transitions and mean field dynamics of the Hamiltonian in Eq.(\ref{eqn:ham}) were studied in Ref.\cite{fu}. 
The Schwinger boson mapping used in Section \ref{sec:tunnel} implies that the pair tunneling term can be identified with two-axis twisting of a spin-$N/2$ particle \cite{kitagawa,norireview} via the equalities $a_{0}^{\dagger 2}a_{1}^{2} + a_{1}^{\dagger 2}a_{0}^{2} = J_{+}^{2} + J_{-}^{2} = 2(J_{x}^{2} - J_{y}^{2})$. The two-axis twisting nonlinearity also appears in the bosonization of the Lipkin-Meshkov-Glick model \cite{kleinboson,parislmg,oberthalerlipkin,opatrnylipkin}. In the context of quantum metrology, it was shown that for a two-mode system that is initialized in a spin coherent state, time evolution generated by two-axis twisting produces states that are useful for magnetometry with precision exceeding the standard quantum limit even in the presence of local non-Markovian dephasing \cite{munroknotttwoaxistwist}.  Unfortunately, for $N>22$, there is no analytical solution of two-axis twisting dynamics \cite{bhattacharyatwisting}. In this Section, we provide a family of variational probe states that allows an experimenter to measure the strength of two-axis twisting with nearly optimal precision.

Just as for the other quartic terms of the Hamiltonian Eq.(\ref{eqn:ham}), the optimal $\mathcal{O}(N^{-4})$ scaling of the QCR bound Eq.(\ref{eqn:qcrtime}) for the mean square error of an estimator of two-axis twisting strength arises due to the fact that the spectral radius of this operator scales at $\mathcal{O}(N^{2})$. Specifically, a $\mathcal{O}(N^{2})$ lower bound for the spectral radius is established by noting that $\sqrt{ \langle \psi_{4} \vert (J_{+}^{2} + J_{-}^{2})^{2}\vert \psi_{4} \rangle}$  computed from Eq.(\ref{eqn:cohstatesuperposvar}) is found to be approximately $(N(N-1)(N-2)(N-3)/4)^{1/2}$ (exactly this value for $N\equiv \pm 2 \mod 8$). An upper bound for the spectral radius is established by noting that $\Vert J_{+}^{2} + J_{-}^{2} \Vert \le 2(\Vert J_{x} \Vert^{2} + \Vert J_{y} \Vert^{2}) = N^{2}$. 

\begin{figure}
\includegraphics[scale=0.43]{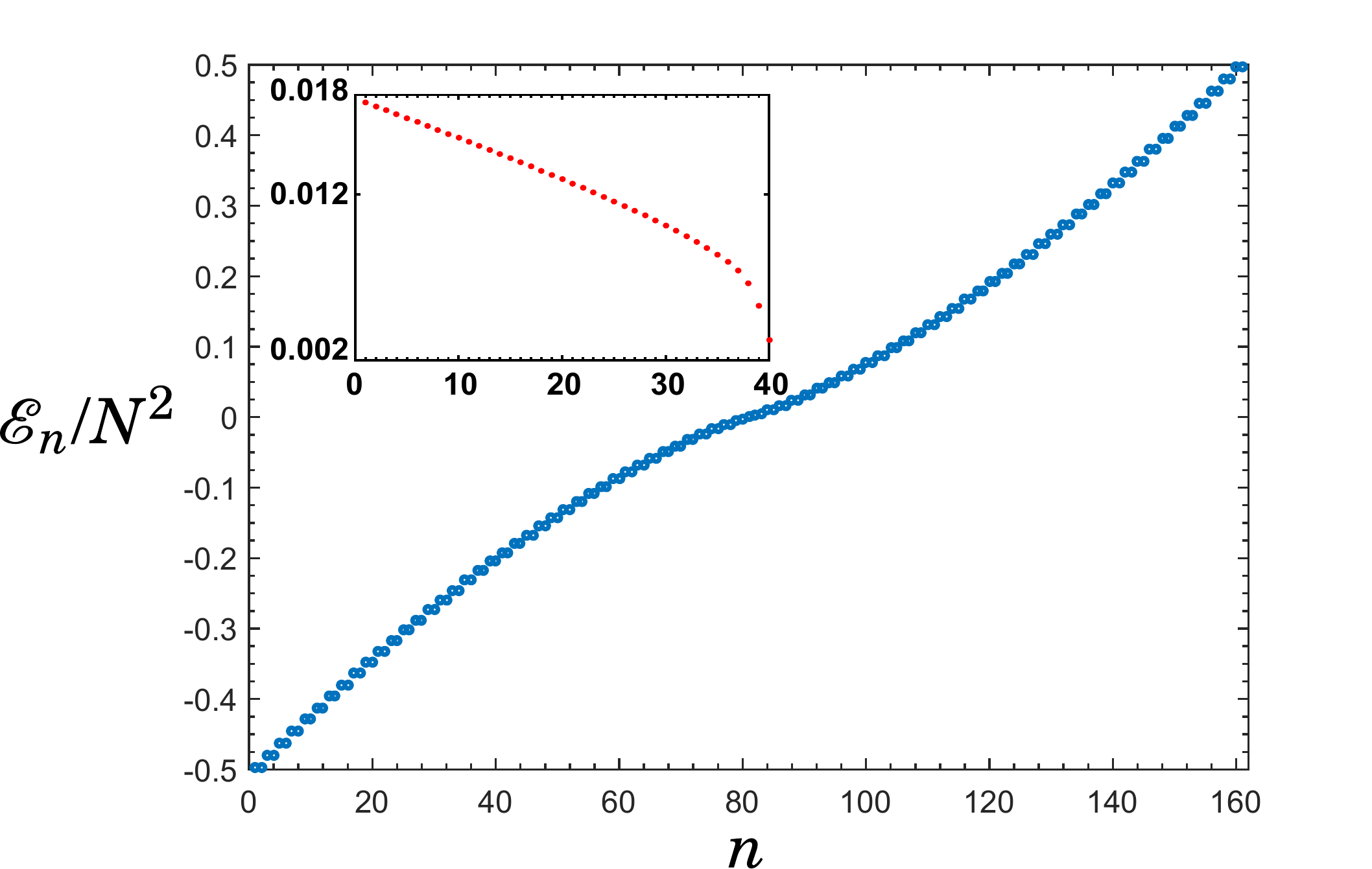}
\caption{Numerically calculated eigenvalues $\mathcal{E}_{n}$ of the Hamiltonian $a_{0}^{\dagger 2}a_{1}^{2} +h.c.$ acting on $N=160$ bosons for $n = 1,\ldots , 161$. The inset shows the gap  $\mathcal{E}_{2n+1}-\mathcal{E}_{2n}$ between pairs of numerical eigenvalues for $n=1,\ldots, 40$. \label{fig:eigenvals}}
\end{figure}

Presently, we focus on the case of $N$ even and present data for the case of $N\equiv 0\mod 4$. We will briefly consider the case of odd $N$ at the end of Section \ref{sec:coherentstates} and in the Appendix. For even $N$, the eigenvalues of $J_{+}^{2} + J_{-}^{2}$ are nondegenerate, but are closely paired  with interpair gap greatly exceeding the intrapair gap (see Fig.\ref{fig:eigenvals}).  We consider the following parametrized states:
\begin{eqnarray}
\ket{\omega_{\pm}(c)}&:=&{1\over \mathcal{N}}\left[ \left( a_{0}^{\dagger 2}+2ica_{0}^{\dagger}a_{1}^{\dagger} - a_{1}^{\dagger 2} \right)^{M} \nonumber \right. \\  & \pm & \left. \left( a_{0}^{\dagger 2}-2ic a_{0}^{\dagger}a_{1}^{\dagger}- a_{1}^{\dagger 2} \right)^{M} \right]\ket{0,0} \label{eqn:superposvar}
\end{eqnarray}
where $c\in \mathbb{R}$, $\mathcal{N}$ is a normalization factor, and $M=N/2$.  Note that for $N$ even, $\ket{\omega_{\pm}(c)}$ maps to $(-1)^{N/2}\ket{\omega_{\pm}(c)}$ under the mode exchange $\ket{0} \leftrightarrow \ket{1}$, i.e., this state is invariant under mode exchange up to a global phase factor. The state $\ket{\omega_{+}(c)}$ ($\ket{\omega_{-}(c)}$) has nonzero amplitude on Dicke basis vectors $\ket{N-k,k}$ for $k$ even ($k$ odd) only. The normalization factor $\mathcal{N}$ can be computed in terms of the hypergeometric distribution, but is not relevant to the present discussion.
\begin{figure}
\includegraphics[scale=0.38]{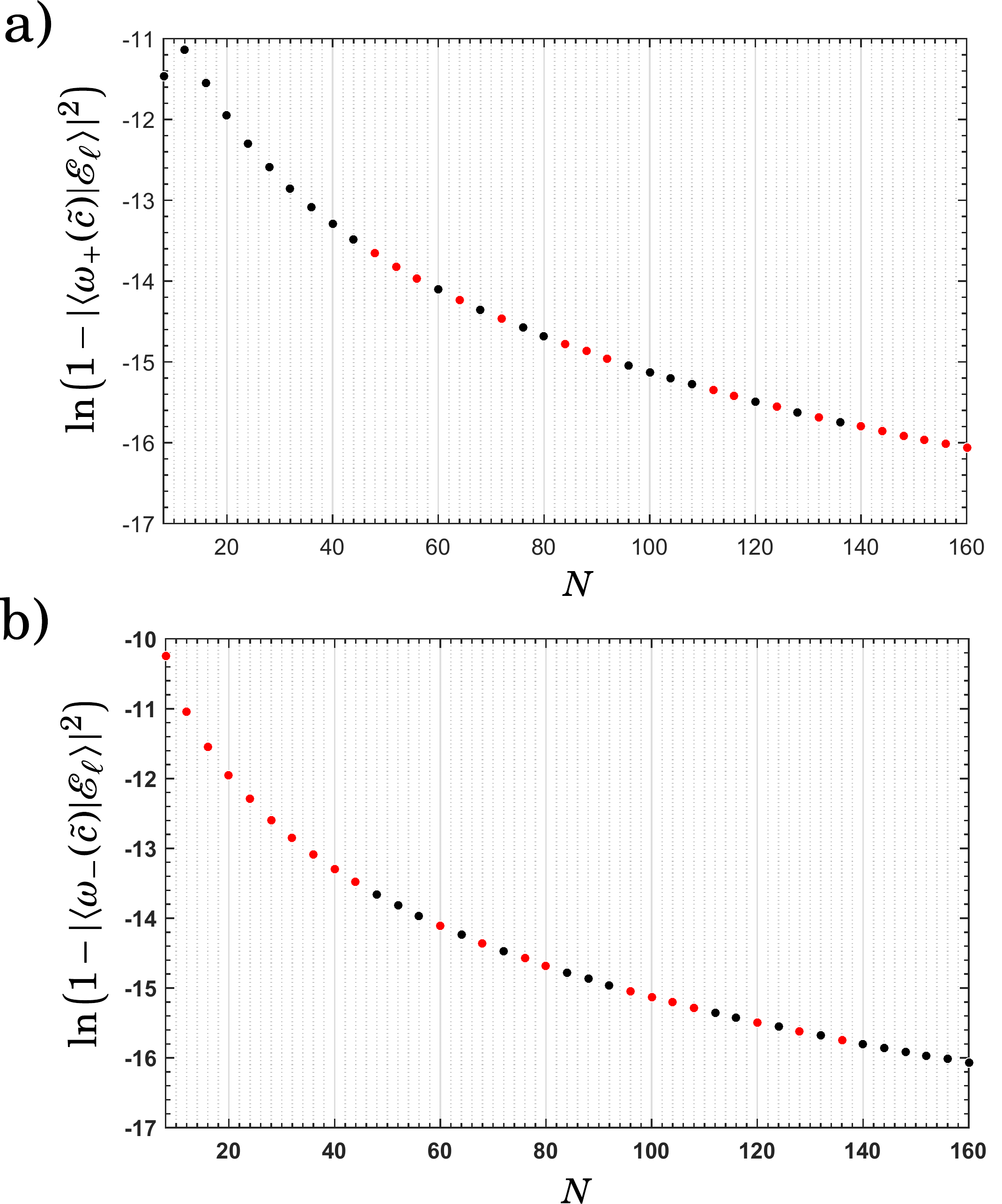}
\caption{a) Semilog plot of the fidelity of the variational state $\ket{\omega_{+}(\tilde{c})}$ with numerical ground state $\ket{\mathcal{E}_{\ell = 1}}$ (black dots) or numerical first excited state $\ket{\mathcal{E}_{\ell = 2}}$  (red dots) for particle number $N$ varying from 8 to 160 in steps of 4. b) The same, except with $\ket{\omega_{-}(\tilde{c})}$.  \label{fig:overlap}}
\end{figure}
For $N=4$, $\ket{\omega_{+}(c)}$ is the exact ground state of $J_{+}^{2}+J_{-}^{2}$ when $c=\left( {\sqrt{3}-1 \over 2}\right)^{1/2}$. For large but finite $N$, the parameter $c$ maximizing the overlap of $\ket{\omega_{\pm}(c)}$ with the ground state of $a_{0}^{\dagger 2}a_{1}^{2}+h.c.$  must be computed numerically. By using the eigenvector consistency conditions Eq.(\ref{eqn:fulldiff}), one can derive an $N$-th order polynomial equation for $c$, the roots of which correspond to exact eigenvectors. We conjecture that for each $N\ge4$ there exists a value $c_{N}$ for which $\ket{\omega_{+}(c_{N})}$ or $\ket{\omega_{-}(c_{N})}$ is the exact ground state. However, a near-perfect variational ground state can be produced by applying only the first two eigenvector consistency conditions of Eq.(\ref{eqn:fulldiff}) to the Dicke state amplitudes of, e.g., $\ket{\omega_{+}(c)}$. For $N$ even, these conditions are $f_{0}C_{2} = \lambda C_{0}$ and $f_{0}C_{0}+f_{2}C_{4}=\lambda C_{2}$, where $f_{k}:=\sqrt{(N-k-1)(N-k)(k+1)(k+2)}$ and $C_{k}:=\langle N-k,k \vert \omega_{+}(c) \rangle$. Solving this pair of equations for $c$ and $\lambda$ gives the solution pair $(\tilde{c},\tilde{\lambda})$
\begin{eqnarray}
\tilde{c}&=&\sqrt{{N-3+\sqrt{N^{2}-2N+3} \over 4N-6}} \nonumber \\
\tilde{\lambda} &=& -2N(1+2\tilde{c}^{2}) - 4N^{2}\tilde{c}^{2}.
\end{eqnarray} Note that this solution does not guarantee that $\ket{\omega_{+}(\tilde{c})}$ satisfies the full list of Eq.(\ref{eqn:fulldiff}) in general and so $\ket{\omega_{+}(\tilde{c})}$ is not necessarily an eigenvector. The $N\rightarrow \infty$ asymptotical behavior of this solution is $\tilde{c}\rightarrow 1/\sqrt{2}$ and $\tilde{\lambda} \rightarrow -N^{2}/2$. In Fig.\ref{fig:overlap} a) and b), the log fidelity of the variational state Eq.(\ref{eqn:superposvar}) with either the numerical ground state (black) or numerical first excited state is shown. From these figures, it is clear that when $\ket{\omega_{+}(\tilde{c})}$ ($\ket{\omega_{-}(\tilde{c})}$) exhibits high overlap with the ground state, $\ket{\omega_{-}(\tilde{c})}$ ($\ket{\omega_{+}(\tilde{c})}$) exhibits high overlap with the first excited state, which is nearly degenerate with the ground state (see Fig. 1). The exceptional agreement with the numerical ground states indicate that the states $\ket{\omega_{\pm}(\tilde{c})}$ capture quantitatively the physics of the ground state of pair tunneling dynamics in the thermodynamic limit. A different variational parameter pair $(\tilde{c}',\tilde{\lambda}')$ is obtained by applying the eigenvector consistency conditions $f_{1}C_{3}=\lambda C_{1}$ and $f_{1}C_{1}+f_{3}C_{5}=\lambda C_{3}$ to $\ket{\omega_{-}(c)}$. However, the fidelities of the states $\ket{\omega_{\pm}(\tilde{c} ')}$ with the numerical ground state exhibit the same $N\rightarrow \infty$ asymptotic behavior as shown in Fig. \ref{fig:overlap}.

The chiral symmetry of the pair tunneling Hamiltonian given by $e^{i{\pi\over 2}J_{z}}\left( J_{+}^{2} + J_{-}^{2} \right) e^{-i{\pi\over 2}J_{z}} = -\left( J_{+}^{2} + J_{-}^{2} \right)$ implies the relation $\ket{\lambda_{\text{max}}} = e^{-i{\pi\over 2}J_{z}}\ket{\lambda_{\text{min}}}$ between the nondegenerate ground state $\ket{\lambda_{\text{min}}}$ and the highest energy state  $\ket{\lambda_{\text{max}}}$. A short calculation shows that \begin{eqnarray}
e^{-i{\pi\over 2}J_{z}}\ket{\omega_{\pm}(c)} &=& {i^{N}\over \mathcal{N}}\left[ \left( a_{0}^{\dagger 2}+2ca_{0}^{\dagger}a_{1}^{\dagger} + a_{0}^{\dagger 2} \right)^{M} \nonumber \right. \\  & \pm & \left. \left( a_{0}^{\dagger 2}-2c a_{0}^{\dagger}a_{1}^{\dagger}+ a_{0}^{\dagger 2} \right)^{M} \right]\ket{0,0} .
\end{eqnarray}
The family of variational probe states for near-optimal estimation of $A_{2}$ can now be obtained directly from Eq.(\ref{eqn:variationalgen}). For $N \equiv 0 \mod 4$, the identity $\langle \omega_{-}(c) \vert e^{-i{\pi \over 2}J_{z}} \vert \omega_{-}(c) \rangle = 0$ holds. Consequently, for $N$ such that $N\equiv 0 \mod 4$ and such that $\ket{\omega_{-}(\tilde{c})}$ exhibits higher fidelity with the numerical ground state than does $\ket{\omega_{+}(\tilde{c})}$, the near-optimal family consists of states $\ket{\Psi_{A_{2}}}$ of the form: \begin{equation}\ket{\Psi_{A_{2}}} = {1\over \sqrt{2}}\left(\ket{\omega_{-}(\tilde{c})} +e^{i\eta}e^{-i{\pi\over 2}J_{z}}\ket{\omega_{-}(\tilde{c})}\right) . \label{eqn:pairtunnearopt}\end{equation} When $\ket{\omega_{+}(\tilde{c})}$ exhibits higher fidelity with the numerical ground state than does $\ket{\omega_{-}(\tilde{c})}$, the value of $w$ in Eq.(\ref{eqn:variationalgen}) is nonzero and can be taken into account when defining the family of near-optimal variational probe states.

 For the data shown in Fig.\ref{fig:overlap}, the largest difference of QFI $4\langle (\Delta A)^{2} \rangle_{\ket{\psi_{\text{true}}}}-4\langle (\Delta A)^{2} \rangle_{\ket{\psi_{\text{var}}}}$ (with $\ket{\psi_{\text{var}}}$ taken to be the family of superpositions in Eq.(\ref{eqn:pairtunnearopt}) and $\ket{\psi_{\text{true}}}$ taken to be the numerical result for the optimal family of superpositions) has a value $\approx 9.2258$ for $N=160$. Using the QCR bound, one thus finds that in a system of 160 resource particles, the cost of using the analytically determined family  Eq.(\ref{eqn:pairtunnearopt}) instead of the optimal family of states is less than losing a single resource particle, in the sense that the maximal QFI for $N=159$ is far lower than the submaximal QFI obtained in the family of states of Eq.(\ref{eqn:pairtunnearopt}) for $N=160$. However, because of the nonlinearity of the interaction $a_{0}^{\dagger 2}a_{1}^{2} + h.c.$ in the bosonic operators, this is not surprising. In Sec. \ref{sec:coherentstates}, we discuss the operational implications of using a lower fidelity variational probe for metrology of $A_{2}$.

\subsection{\label{sec:coherentstates}Superpositions of spin coherent states: even and odd N}
\begin{figure}
\includegraphics[scale=0.38]{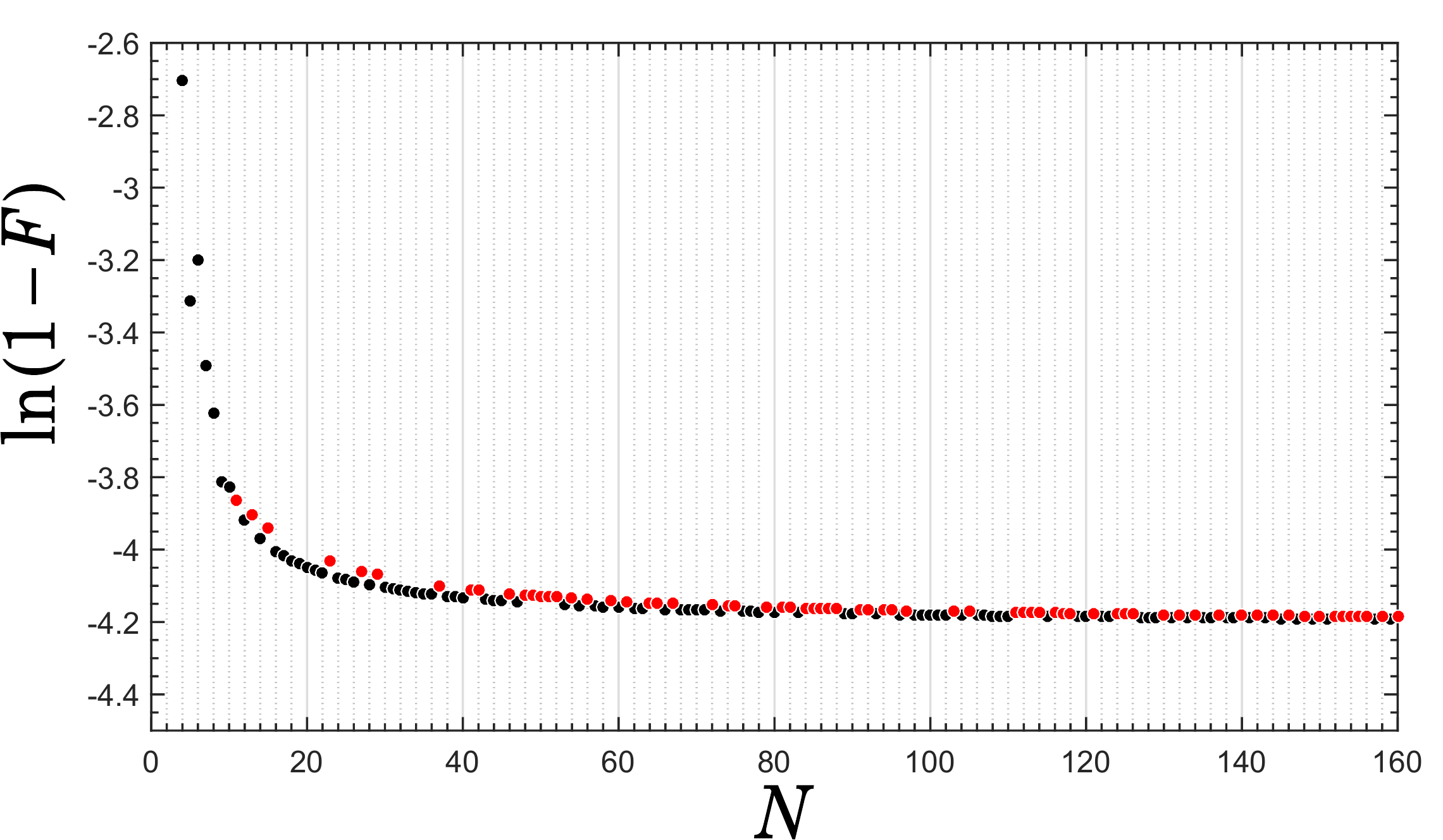}
\caption{Semilog plot of the infidelity $1-F(\ket{A},\ket{B})=1-\vert \langle A \vert B\rangle \vert^{2}$ for $\ket{A} := 1/\sqrt{2}\left( \ket{\zeta=i} \pm \ket{\zeta = -i} \right)$ and  $\ket{B}:=\ket{\mathcal{E}_{\ell = 1}}$ the numerical ground state. The black (red) dots corrspond to values of $N$ where the $+$ ($-$) sign of $\ket{A}$ gives better agreement with $\ket{\mathcal{E}_{\ell=1}}$.  \label{fig:cohstateoverlap}}
\end{figure}

It is useful to consider the decrease in metrological performance incurred by taking $c=1$ in Eq.(\ref{eqn:pairtunnearopt}). In this limit, these states become the state in Eq.(\ref{eqn:tunopt}) with $\vec{n}=(0,1,0)$. The fidelity of the superposition states $1/\sqrt{2}\left( \ket{\zeta=i}\pm\ket{\zeta=-i} \right)$ are shown in Fig.\ref{fig:cohstateoverlap}. To explicitly calculate the variance of $J^{2}_{+} + J_{-}^{2}$ in the superposition $\ket{\psi_{4}}\propto \ket{\zeta=i}+\ket{\zeta=-i} + \ket{\zeta = 1} + \ket{\zeta = -1}$, we make use of the following formulas for coherent state matrix elements of the raising and lowering operators:
\begin{eqnarray}
\langle \zeta ' \vert J_{-}^{m}J_{+}^{n}\vert \zeta \rangle &=&{ \del^{m}_{\overline{\zeta'}}\del^{n}_{\zeta}\left(1+\overline{\zeta'}\zeta\right)^{N} \over  \left((1+\vert \zeta ' \vert^{2})(1+\vert \zeta  \vert^{2}) \right)^{{N\over 2}}}
\nonumber \\
\langle \zeta ' \vert J_{+}^{m}J_{-}^{n}\vert \zeta \rangle &=&{ \del^{m}_{\overline{\zeta'}^{-1}}\del^{n}_{\zeta^{-1}}\left(1+(\overline{\zeta'}\zeta)^{-1}\right)^{N} \over  \left((1+\vert \zeta ' \vert^{2})(1+\vert \zeta  \vert^{2}) \right)^{{N\over 2}}} .
\end{eqnarray}
The variance of pair tunneling, $\langle \psi_{4} \vert (\Delta (J_{+}^{2}+J_{-}^{2}))^{2} \vert \psi_{4} \rangle$,  is given by
\begin{eqnarray}
&&{1\over 1+2^{-{N\over 2}+1}\cos (N\pi /4)}\left[ {N(N-1)(N-2)(N-3)\over 4}  \right. \nonumber \\ && \left. - 2^{-{N\over 2}+1}(N-1)(N-2)(N-3)\cos {(N-4)\pi \over 4} \right].
\label{eqn:cohstatesuperposvar}
\end{eqnarray}

Although it is clear from this expression that the quantum Fisher information scales as $\mathcal{O}(N^{4})$, one expects from comparing the fidelity data in Fig.\ref{fig:cohstateoverlap} to the data in Fig.\ref{fig:overlap} that the state $\ket{\psi_{4}}$ is a worse probe for estimation of $A_{2}$ than the family of states in Eq.(\ref{eqn:pairtunnearopt}).
To compare the metrological usefulness of $\ket{\psi_{4}}$ with that of the family of variational probe states derived in  Eq.(\ref{eqn:pairtunnearopt}), we first note that the difference in QFI $4\langle (\Delta A)^{2} \rangle_{\ket{\psi_{\text{true}}}}-4\langle (\Delta A)^{2} \rangle_{\ket{\psi_{\text{var}}}}$ between the numerical ground state $\ket{\psi_{\text{true}}}$ and the state $\ket{\psi_{\text{var}}}=\ket{\psi_{4}}$ has a value  $\approx 1.3\times 10^{6}$ when  $N=160$.  This difference still warrants the use of $\ket{\psi_{4}}$ as a probe for metrology of $A_{2}$ before settling for the exactly optimal state for $N=159$ resource particles. The operational inefficiency of using a poor variational state as a probe in a quantum metrology protocol can be shown by writing the QCR bound for a variational probe state $\ket{\psi_{A}}$ as $1/( \nu_{A}(\mathcal{F}_\text{max}-x_{A}))$, where $\mathcal{F}_{\text{max}}-x_{A}$ is the QFI of $\ket{\psi_{A}}$ written as a deviation from the maximum possible QFI and $\nu_{A}$ symbolizes the number of runs of an experiment estimating the relevant real parameter. If another nonoptimal probe state gives a QCR bound of $ 1/(\nu_{B}(\mathcal{F}_{\max}- x_{B} ))$ with $x_{B}>x_{A}$, then $\nu_{B}/\nu_{A} = {1-{x_{A}\over \mathcal{F}_{\text{max}}  } \over 1-{x_{B}\over \mathcal{F}_{\text{max}}  }}$ for the QCR bounds to be equal. For $N=160$, we find the ratio $\nu_{B} /\nu_{A} \approx 1.01$ for the number of experiments that must be run by an experimenter using $\ket{\psi_{B}} = \ket{\psi_{4}}$ to the number of experiments that must be run by an experimenter using $\ket{\psi_{A}}= \ket{\omega_{-}(\tilde{c})} + e^{-i{\pi/2}J_{z}}\ket{\omega_{-}(\tilde{c})}$ as the probe state.

We note that for $N$ odd, the two-axis twisting Hamiltonian $J_{+}^{2} +J_{-}^{2}$ exhibits twofold degenerate eigenvectors (see Appendix \ref{sec:appa}). This leads to a richer family of near-optimal variational states. In this case, the ground state subspace is spanned by an even state $\ket{\lambda^{(e)}_{\text{min}}}$ and an odd state $\ket{\lambda^{(o)}_{\text{min}}}$ having, respectively, nonzero amplitudes on Dicke states $\ket{N-k,k}$ with $k$ even and odd. The eigenvectors corresponding to the largest eigenvalue are $\ket{\lambda^{(e)}_{\text{max}}} := e^{-i{\pi \over 2}J_{z}}\ket{\lambda_{\text{min}}^{(e)}}$ (analogously for $\ket{\lambda^{(o)}_{\text{max}}}$). Defining $\ket{\Psi_{\text{min}(\text{max})}(\theta,\varphi)} := (1/ \sqrt{2})\left( \cos {\theta \over 2} \ket{\psi_{\text{min}}^{(e)}} +\sin {\theta \over 2}e^{i\varphi} \ket{\psi_{\text{min}}^{(o)}}\right) $, the maximal variance of $J_{+}^{2} + J_{-}^{2}$ is obtained for family of probe states
\begin{equation}
{1\over \sqrt{2}}\left(\ket{\Psi_{\text{min}}(\theta,\varphi)} + e^{i\eta}\ket{\Psi_{\text{max}}(\theta',\varphi')} \right)
\label{eqn:oddNopt}
\end{equation}
parametrized by $S^{2}\times S^{2}\times S^{1}$ with coordinates $(\theta , \varphi)$, $(\theta ' , \varphi ')$, $\eta$.

\subsection{\label{subsection:qptmetro}Single particle tunneling estimation with a ground state probe}

A natural question that arises with regard to quantum estimation of the coupling constants of the Hamiltonian Eq.(\ref{eqn:ham}) is whether the ground state is ever near-optimal for estimation of any one of the coupling constants. Recall from Section \ref{sec:tunnel} that the optimal states for estimation of a single-particle tunneling amplitude are superpositions of antipodal coherent states.  The fact that $\ket{\omega_{\pm}(\tilde{c})}$ introduced in Section \ref{sec:pairtun} are good variational ground states for $a_{0}^{\dagger 2}a_{1}^{2} +h.c.$ and also exhibit a two-component superposition structure with support on either even or odd Dicke states suggests that the true ground state of the two-particle tunneling term may provide a ``natural'' resource for estimation of $A_{1}$.

\begin{table}
\caption{Normalized quantum Fisher information on the path generated by $2J_{x}$ with initial state the numerical ground state $\ket{\lambda_{\text{min}}}$ of $-2\left(J_{x}^{2}-J_{y}^{2}\right)$.\label{table:tab1}}
\begin{tabular}{lc}
  N & $\mathcal{F}(\ket{\lambda_{\text{min}}})/N^{2}$  \\ \hline \hline
  4 & 0.9330  \\
  36 & 0.9965 \\
  68 & 0.9982  \\
  100 & 0.9988  \\
  132 & 0.9991  \\
  160 & 0.9992  \\ \hline \hline
\end{tabular}
\end{table}

To verify this, we display in Table \ref{table:tab1} values of the scaled QFI, $\mathcal{F}(\ket{\lambda_{\min}})/N^{2}=4\langle (\Delta H)^{2} \rangle_{\ket{\lambda_{\text{min}}}} /N^{2}$, corresponding to the Hamiltonian $H=2J_{x}$ and the numerical ground state $\ket{\lambda_{\text{min}}}$ of $A=-2\left( J_{x}^{2}-J_{y}^{2} \right)$ for various $N$. The minus sign multiplying $\left( J_{x}^{2}-J_{y}^{2} \right)$ in the definition of $A$ has the physical consequence that the ground state of $A$ exhibits constructive interference between amplitudes on Dicke states $\ket{N-k,k}$ and $\ket{N-(k\pm 2), k\pm 2}$. This can be achieved in a Bose gas with negative $s$-wave scattering length or by engineering a $\pi \over 2$ phase shift between the $\ket{0}$ and $\ket{1}$ single particle states so that $z^{2} = -1$ in  Eq.(\ref{eqn:identify}). Although a magnitude of $\mathcal{O}(1/N)$ separates the observed QFI from the maximal QFI, the results suggest that by cooling and adiabatically  tuning the two-mode system governed by the tunneling Hamiltonian $H_{\text{tun}}:= 2A_{1}J_{x} - 2A_{2}\left( J_{x}^{2}-J_{y}^{2} \right)$  through a quantum phase transition into the parameter regime $A_{1}$, $A_{2}>0$ and $A_{2}\gg A_{1}$  provides a method of generating near-optimal states for estimation of the single particle tunneling amplitude.

\section{\label{sec:weighttun}Number-weighted single particle tunneling} 

We now consider the number-weighted tunneling term  $(T_{0}a_{0}^{\dagger}a_{0}+T_{1}a_{1}^{\dagger}a_{1})a_{0}^{\dagger}a_{1} + h.c.$  in Eq.(\ref{eqn:ham}). In the case that $T_{0}=T_{1}$ holds identically, the number-weighted single particle tunneling term becomes $T_{0} \left( \sum_{k=0}^{1} a_{k}^{\dagger }a_{k} \right) \left( a_{0}^{\dagger}a_{1} +a_{1}^{\dagger}a_{0} \right)$ which, when restricted to a system of $N$ bosons, simply renormalizes $A_{1}$ by $A_{1}\mapsto A_{1} + T_{0}N$. The single particle tunneling amplitude can then be estimated using the optimal probe state of Section \ref{sec:tunnel}. Because the matrix representation of the number-weighted tunneling terms is tridiagonal and symmetric, the spectrum is nondegenerate for all $N$.

In the case that $T_{0}\neq T_{1}$, a variational method is again required for the derivation of near-optimal families of states for quantum estimation of $T_{0}$ and $T_{1}$. In this section we take $T_{1}=0$ and focus on estimation of $T_{0}$ (the case of $T_{0}=0$, $T_{1}\neq 0$ can be treated in the same way). The corresponding number-weighted tunneling term $a_{0}^{\dagger}a_{0}a_{0}^{\dagger}a_{1} + h.c. = \left( N-J_{z} \right) J_{-} + J_{+}\left( N-J_{z} \right)$ exhibits a chiral symmetry under the rotation $e^{i\pi J_{z}}$. Therefore, a variational probe state of the form appearing in Eq.(\ref{eqn:variationalgen}) can be constructed by taking for $\ket{\psi_{\text{min}}}$ a variational ground state of the number-weighted tunneling term and taking for $\ket{\psi_{\text{max}}}$ the rotation of $\ket{\psi_{\text{min}}}$ by $e^{i\pi J_{z}}$. As was found for the other quartic terms in Eq.(\ref{eqn:ham}), the maximal QFI appearing in the QCR bound for estimation of $T_{0}$ exhibits $\mathcal{O}(N^{4})$ scaling. 

\begin{figure}
\includegraphics[scale=0.43]{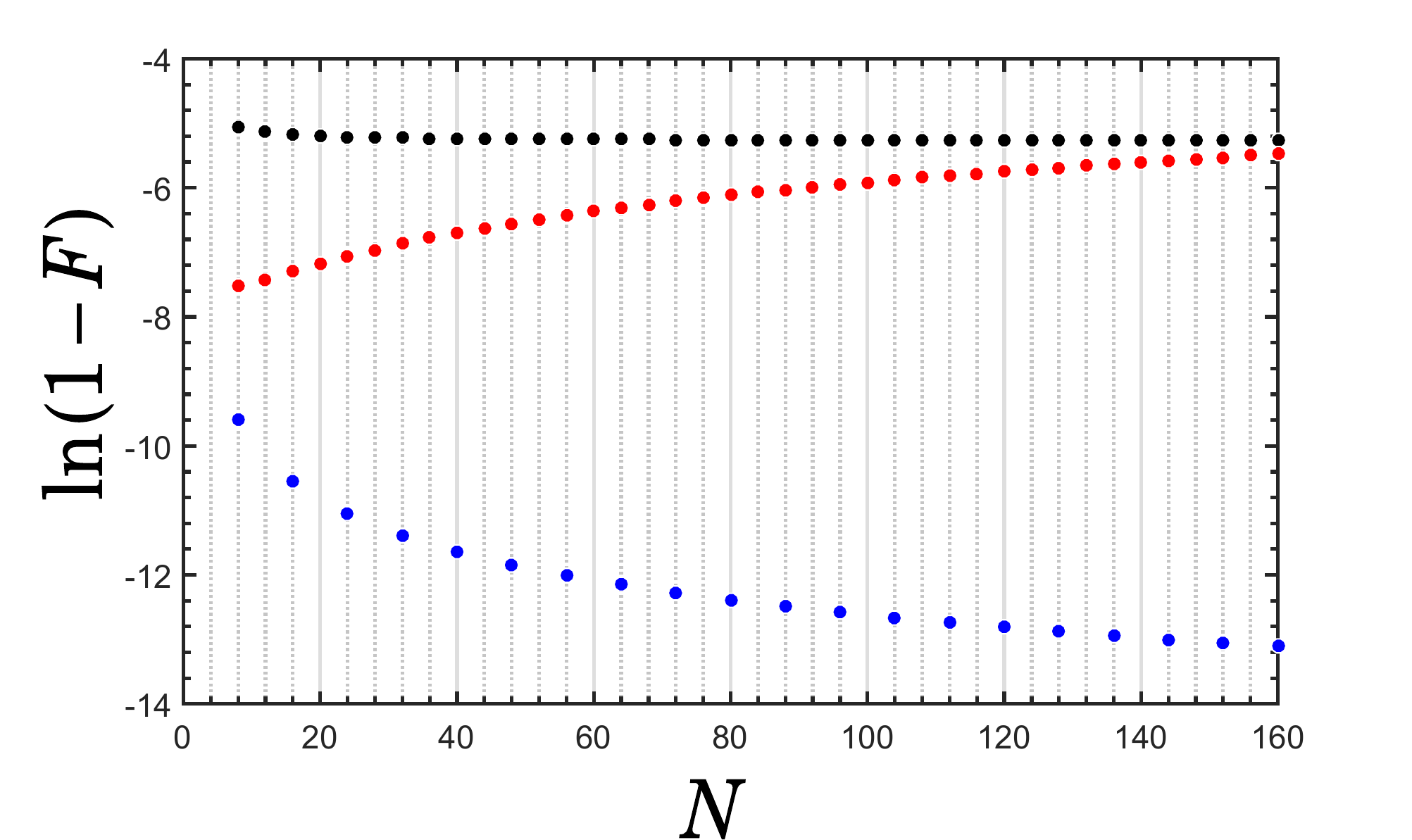}
\caption{Semilog plot of the infidelity $1-F(\ket{A},\ket{B})=1-\vert \langle A \vert B\rangle \vert^{2}$ for $\ket{A}$ a variational ground state,  $\ket{B}:=\ket{\mathcal{E}_{\ell = 1}}$ is the numerical ground state, and $N=8,12,\ldots , 160$. The black dots represent $\ket{A} :=  \ket{\zeta} $ where $\zeta$ extremizes Eq.(\ref{eqn:oddexpect}); the red dots represent $\ket{A} := \ket{\Xi(\tilde{w},\tilde{z})} \propto \left( a_{0}^{\dagger 2}+2\tilde{w}a_{0}^{\dagger}a_{1}^{\dagger} +\tilde{z}^{2}a_{1}^{\dagger 2} \right)^{M} \ket{0,0}$ with $(\tilde{w},\tilde{z})$ determined by the $k=0$ and $k=1$ equations in Eq.(\ref{eqn:nweightedconsist}); the blue dots show $\ket{A} := \ket{\Xi( w_{0},z_{0})}$ with $(w_{0},z_{0})$ determined by numerical optimization. \label{fig:cohstatevspairvar}}
\end{figure}

Because the ground state of any single particle tunneling term is a spin-$N/2$ coherent state, we first consider the best coherent state approximation to the ground state of number-weighted single particle tunneling. The expected value of the number-weighted tunneling term in a coherent state $\ket{\zeta}$ with $\zeta \in \mathbb{R}$ is given by
\begin{equation}
\langle \zeta \vert a_{0}^{\dagger}a_{0} a_{0}^{\dagger}a_{1} +h.c.\vert \zeta \rangle = {2N(N-1)\zeta \over (1+\zeta^{2})^{2}}+{2N\zeta \over 1+\zeta^{2}} 
\label{eqn:oddexpect}
\end{equation}
which is an odd function of $\zeta$. The black circles in Fig. \ref{fig:cohstatevspairvar} show the infidelity of the best coherent state approximation to the numerical ground state, i.e., the coherent state $\ket{\zeta_{0}}$ with $\zeta_{0}$ minimizing Eq.(\ref{eqn:oddexpect}). In comparison to Fig. \ref{fig:cohstateoverlap}, it is clear that the ground state of $a_{0}^{\dagger}a_{0}a_{0}^{\dagger}a_{1}+h.c.$ is better approximated by a coherent state than the ground state of $a_{0}^{\dagger 2}a_{1}^{2} + h.c.$ is approximated by a superposition of two coherent states. For $N \rightarrow \infty$, $\zeta_{0} \rightarrow \pm 1/\sqrt{3}$ are the coherent state parameters extremizing the expectation value. Using the chiral symmetry $e^{i\pi J_{z}}$ and Eq.(\ref{eqn:variationalgen}), a family of variational superposition states for estimation of $T_{0}$ can be obtained. For a state $\ket{\Psi_{T_{0}}}$ of this family, it follows from Eq.(\ref{eqn:oddexpect}) and the fact that $\langle -\zeta \vert a_{0}^{\dagger}a_{0} a_{0}^{\dagger}a_{1} +h.c.\vert \zeta \rangle = 0$ that $\langle \Psi_{T_{0}} \vert  a_{0}^{\dagger}a_{0} a_{0}^{\dagger}a_{1} +h.c. \vert \Psi_{T_{0}} \rangle = 0$, which is a necessary condition on an optimal probe state. However, by identifying a variational ground state of $a_{0}^{\dagger}a_{0}a_{0}^{\dagger}a_{1}+h.c.$ that exhibits higher fidelity with the true ground state as compared to any coherent state, Eq.(\ref{eqn:variationalgen}) may be used to construct a class of superposition probe states with a lower QCR bound for estimation of $T_{0}$ as compared to $\ket{\Psi_{T_{0}}}$.

We can improve on the ground state fidelity achieved by a spin coherent state by making use of the following bivariational pair state :
\begin{equation}
\ket{\Xi(w,z)} \propto \left( a_{0}^{\dagger 2}+2wa_{0}^{\dagger}a_{1}^{\dagger} +z^{2}a_{1}^{\dagger 2} \right)^{M} \ket{0,0} ,
\end{equation} with $M=N/2$. A similar variational state was introduced in Ref.\cite{kawaguchi} to analyze the dynamical instability of the polar phase of a spin-1 Bose gas to formation of a gas of Goldstone magnons. An ansatz for the parameters $w$ and $z$ can be obtained by using the  numerical ground state energy $\lambda_{0}$ in the following necessary and sufficient condition for a state $\sum_{k=0}^{N}C_{k}\ket{N-k,k}$ to be an eigenvector with eigenvalue $\lambda$:
\begin{eqnarray}
&{}&(N-k)\sqrt{(N-k)(k+1)}C_{k+1} \nonumber \\ &+&(N-k+1)\sqrt{(N-k+1)k}C_{k-1} = \lambda C_{k}
\label{eqn:nweightedconsist}
\end{eqnarray}
where $C_{N+1}$ and $C_{-1}$ are defined to be 0.

A solution pair $(\tilde{w},\tilde{z})$ can be obtained from the two equations corresponding to $k=0$ and $k=1$ in Eq.(\ref{eqn:nweightedconsist}). For small and intermediate values of $N$, a large increase in the fidelity  of $ \ket{\Xi(\tilde{w},\tilde{z})}$  with the numerical ground state $\ket{\lambda_{\text{min}}}$ (Fig.\ref{fig:cohstatevspairvar}, red circles) is observed as compared to the fidelity of $\ket{\zeta_{0}}$ with $\ket{\lambda_{\text{min}}}$ (Fig.\ref{fig:cohstatevspairvar}, black circles). However, unlike the analogous approach in Section \ref{sec:pairtun}, the application of the eigenvector consistency condition Eq.(\ref{eqn:nweightedconsist}) to the present variational state results in a state that exhibits monotonically increasing infidelity with the numerical ground state (red circles, Fig.\ref{fig:cohstatevspairvar}). To verify that the state $\ket{\Xi(w,z)}$ is a variational ground state for number-weighted tunneling that exhibits monotonically decreasing infidelity with the numerical ground state as $N$ increases, we numerically minimize the function $\log\left( 1-\vert \langle \Xi (w,z) \vert \lambda_{\text{min}} \rangle \vert^{2} \right)$ over $w$ and $z$ using the \texttt{fminsearch} function in MATLAB (blue circles, Fig.\ref{fig:cohstatevspairvar}) for $N=8,16,\ldots , 160$. Calling the minimizing parameters $(w_{0},z_{0})$ and noting that the unitary operator $e^{i\pi J_{z}}$ implements the chiral symmetry of $a_{0}^{\dagger}a_{0}a_{0}^{\dagger}a_{1}+h.c.$, the states $\ket{\Xi(w_{0},z_{0})}$ and $e^{i\pi J_{z}}\ket{\Xi(w_{0},z_{0})}$ can be used as the variational ground state $\ket{\psi_{\text{min}}}$ and variational highest energy state $\ket{\psi_{\text{max}}}$, respectively, in Eq.(\ref{eqn:variationalgen}).

It should be noted that if $T_{0}$ and $T_{1}$ in Eq.(\ref{eqn:identify}) are real numbers, the number-weighted tunneling contribution to the Hamiltonian of the weakly interacting Bose gas takes the form $\sum_{j=0}^{1}\left( T_{j}a_{j}^{\dagger}a_{j}\left( a_{0}^{\dagger}a_{1} + a_{1}^{\dagger}a_{0} \right) + h.c. \right)$. Restricting to the problem of optimal estimation of $T_{0}$ without loss of generality, it is clear that one may still utilize the near-optimal probe states $\ket{\Xi(w_{0},z_{0})}$ without changing the QCR bound due to the fact that the operators $a_{0}^{\dagger}a_{0} a_{0}^{\dagger}a_{1} + h.c.$ and $a_{0}^{\dagger}a_{0} a_{1}^{\dagger}a_{0} + h.c.$ differ by an element of $\mathfrak{su}(2)$ which renormalizes the single particle tunneling amplitude. This follows from the commutation relations $[a_{0},a_{0}^{\dagger}a_{0}]=a_{0}$, $[a_{0}^{\dagger},a_{0}^{\dagger}a_{0}]= - a_{0}^{\dagger}$ that, unlike the canonical commutation relations, are valid on $S(\mathbb{C}^{2})^{\otimes N}$ as well as on the Hilbert space of a quantum harmonic oscillator.
 
\section{Conclusion}

By expanding the Hamiltonian of the weakly-interacting Bose gas over two orthogonal single particle states, we have identified the minimal set of parameters defining the real dynamics of this system. For interactions diagonal in the Dicke basis or for single particle tunneling processes, the $\mathfrak{su}(2)$ symmetry allows for simple identification of the optimal states for (separate) quantum estimation of the real coupling constants or tunneling amplitudes, respectively. In the case of tunneling amplitude estimation or relative phase estimation, the optimal states are superpositions of antipodal coherent states, which leads to the conclusion that a value of unity for the degree of fragmentation $F_{D}$ is necessary for an optimal probe state of these parameters. In the absence of an experimental method for engineering superpositions of antipodal coherent states, near-optimal probe states for single particle tunneling amplitudes can be ``naturally'' obtained by first tuning the coherent pair tunneling process to energetically dominate over single particle tunneling and subsequently allowing relaxation to the ground state.

For Dicke state non-diagonal interactions that do not possess analytical solutions for all $N$, variational methods can be applied to identify near-optimal probe states for estimation of the corresponding coupling constants. Interactions that possess chiral symmetry allow a near-optimal family of variational probe states to be derived from high fidelity ground states. For the case of pair tunneling and number-weighted tunneling processes, the parametrized variational ground states $\ket{\omega_{\pm}(c)}$ and $\ket{\Xi(w,z)}$, respectively, allow (via Eq.(\ref{eqn:variationalgen}) ) to define near-optimal variational probe families that outperform families composed superpositions of few coherent states with the appropriate symmetry. We have also quantified the reduction of operational efficiency incurred by using a suboptimal family of variational probe states in a quantum metrology protocol. 

We expect that the optimal and near-optimal probe states for quantum estimation of the two-mode weakly interacting Bose gas will inform quantum technologies that exploit ultracold atomic Bose gases for quantum metrology at precisions beyond the limits imposed by the use of product states. Two pressing problems are present challenges for the implementation of quantum metrology protocols in ultracold Bose gases: 1) identification of optimal or near-optimal states for multiparameter estimation of Eq.(\ref{eqn:ham}), and, 2) experimental generation of optimal and near-optimal probe states and measurements.
\bibliography{nbosons.bib}

\appendix

\section{Degeneracy of eigenvectors of $J_{+}^{2}+J_{-}^{2}$ for odd $N$\label{sec:appa}}

Calculation of the spectrum and eigenvectors of $ J_{+}^{2} + J_{-}^{2}$ via a traditional method, e.g., extremizing $\langle J_{+}^{2} + J_{-}^{2}\rangle$ over pure spin-$N/2$ states, leads to the second order difference equation \begin{eqnarray}f_{0}C_{2} &=& \lambda C_{0} \nonumber \\ f_{1}C_{3} &=& \lambda C_{1} \nonumber \\ f_{k-2}C_{k-2} + f_{k}C_{k+2} &=& \lambda C_{k} , \; k\in\lbrace 2,\ldots , N-2\rbrace \nonumber \\ f_{N-3}C_{N-3} &=& \lambda C_{N-1} \nonumber \\ f_{N-2}C_{N-2} &=& \lambda C_{N}    \label{eqn:fulldiff}\end{eqnarray} for the amplitudes $C_{k}$ of the state $\ket{\psi_{\lambda}} = \sum_{k=0}^{N}C_{k}\ket{N-k,k}$ satisfying $ (J_{+}^{2} + J_{-}^{2})\ket{\psi_{\lambda}} = \lambda \ket{\psi_{\lambda}}$, where $f_{k} :=\sqrt{(N-k-1)(N-k)(k+1)(k+2)}$. For $N$ even, the eigenvalues of $J_{+}^{2}+J_{-}^{2}$ are nondegenerate while for $N$ odd, the eigenvalues have multiplicity 2. The proof of the degeneracy for $N$ odd proceeds by taking $\lbrace C_{k} \rbrace_{k=0}^{N}$ to be a solution of Eq.(\ref{eqn:fulldiff}) with eigenvalue $\lambda$ and noting that $f_{N-k} = f_{k-2}$. It follows from Eq.(\ref{eqn:fulldiff}) that  
\begin{eqnarray}  f_{0}C_{N-2} &=& \lambda C_{N}\nonumber \\ f_{3}C_{N-3} &=& \lambda C_{N-1} \nonumber \\  f_{k-2}C_{N-k+2} + f_{k}C_{N-k-2}  &=& \lambda C_{N-k} , \; k\in\lbrace 2,\ldots , N-2\rbrace \nonumber \\    f_{N-2}C_{2} &=& \lambda C_{0} \nonumber \\ f_{N-3}C_{3} &=& \lambda C_{1}. \label{eqn:transfdiff}\end{eqnarray} 
This equation allows to construct a solution $\lbrace C'_{k} \rbrace_{k=0}^{N}$ of Eq.(\ref{eqn:fulldiff}) with the same value of $\lambda$, where $C_{k}' = \overline{C_{N-k}}$ if $k$ odd and $C_{k}' = -\overline{C_{N-k}}$ if $k$ even. The solution defined by $\lbrace C'_{k} \rbrace_{k=0}^{N}$ is orthogonal to the solution $\lbrace C_{k} \rbrace_{k=0}^{N}$ because \begin{eqnarray}\sum_{k=0}^{N}\overline{C_{k}}C_{k}' &=& \sum_{k=0}^{(N-1)/2}\overline{C_{k}}C_{k}' +\overline{C_{N-k}}C_{N-k}' \nonumber \\ &=& (-\overline{C_{0}}\,\overline{C_{N}} +\overline{C_{N}}\,\overline{C_{0}}) + (\overline{C_{1}}\,\overline{C_{N-1}} - \overline{C_{N-1}}\,\overline{C_{1}})+\ldots \nonumber \\ &=&0. \end{eqnarray}

\end{document}